\documentclass[prmaterials,psfig,pstricks,twocolumn,preprintnumbers,amsmath,amssymb,superscriptaddress]{revtex4}

\usepackage{graphicx}          
\usepackage{dcolumn}           
\usepackage{bm}                

\usepackage{array,booktabs,lmodern}
\usepackage[scaled=1.00]{helvet}
\usepackage[final]{listings,microtype}
\usepackage[T1]{fontenc}
\usepackage[version=3]{mhchem} 
\usepackage{siunitx}
\usepackage[english]{babel}
\usepackage{float}
\usepackage{natbib}
\usepackage{setspace}
\usepackage{xkeyval}
\usepackage{natmove}
\usepackage{titlesec}

\usepackage{xfrac}
\usepackage{relsize}

\usepackage{braket}
\usepackage{amssymb}
\defcitealias{chang20183d}{[Chang et al., Science 360, 778-783 (2018)]}

\usepackage{hyperref}
\hypersetup{
    colorlinks=true,
    urlcolor= blue,
    citecolor=blue,
    linkcolor= blue,
    bookmarksopen=false,
    }

\begin{document}

\preprint{Page}

\title{Microscopic origin of the excellent thermoelectric performance in n-doped SnSe} 

\author{Anderson S. Chaves}
\affiliation{Gleb Wataghin Institute of Physics, University of Campinas, PO Box 13083-859, Campinas, SP, Brazil}
\author{Daniel T. Larson}
\affiliation{Department of Physics, Harvard University, Cambridge, Massachusetts, 02138, USA}
\author{Efthimios Kaxiras}
\affiliation{Department of Physics, Harvard University, Cambridge, Massachusetts, 02138, USA}
\affiliation{John A. Paulson School of Engineering and Applied Sciences, Harvard University, Cambridge, Massachusetts, 02138, USA}
\author{Alex Antonelli}
\affiliation{Gleb Wataghin Institute of Physics and Center for Computing in 
Engineering \& Sciences, University of Campinas, PO Box 13083-859, Campinas, SP, Brazil}

\date{\today}

\begin{abstract}

Excellent thermoelectric performance in the out-of-layer n-doped SnSe
has been observed experimentally 
\citetalias{chang20183d}. However, a first-principles investigation
of the dominant scattering mechanisms governing all thermoelectric transport properties is lacking.
In the present work, by applying extensive first-principles calculations of electron-phonon coupling
associated with the calculation of the scattering by ionized impurities, we investigate the
reasons behind the superior figure of merit as well as the enhancement of $zT$ above 600~K
in n-doped out-of-layer SnSe, as compared to p-doped SnSe with similar carrier densities.
For the n-doped case, the relaxation time is dominated by ionized impurity scattering and
increases with temperature, a feature that maintains the power factor at high values at
higher temperatures and simultaneously
causes the carrier thermal conductivity at zero electric current ($\kappa_\mathrm{el}$)
to decrease faster for higher temperatures, leading to an ultrahigh-$zT = 3.1$ at 807~K.
We rationalize the roles played by $\kappa_\mathrm{el}$ and $\kappa^{0}$ (the thermal conductivity due
to carrier transport under isoelectrochemical conditions) in the determination of $zT$.
Our results show the ratio between $\kappa^{0}$ and the lattice thermal conductivity
indeed corresponds to the upper limit for $zT$, whereas the difference
between calculated $zT$ and the upper limit
is proportional to $\kappa_\mathrm{el}$.

\end{abstract}


\maketitle

\section{Introduction}

New materials for energy harvesting applications 
are necessary for reducing greenhouse gas emissions. Thermoelectric (TE) materials that can harvest waste heat 
from traditional nuclear or coal power plants\cite{sootsman2009new,bell2008cooling}
represent a source of cleaner electric power.\cite{disalvo1999thermoelectric,he2017advances} 
However, widespread deployment will require increases in efficiency to compete with other forms of power generation.\cite{disalvo1999thermoelectric,vining2009inconvenient,he2017advances} 
The efficiency of a TE material is characterized by the 
dimensionless figure of merit, $zT = \sigma S^2 T/\kappa_\mathrm{tot}$, 
where $\sigma$ is the electrical conductivity, $S$ is the Seebeck coefficient, $T$ 
is the absolute temperature, and $\kappa_\mathrm{tot} = \kappa_\mathrm{latt} + \kappa_\mathrm{el}$ 
is the total thermal conductivity composed of lattice ($\kappa_\mathrm{latt}$) and 
carrier ($\kappa_\mathrm{el}$) contributions. 

The search for high-$zT$ materials is ongoing\cite{he2017advances}. 
The most common strategies for the optimization of $zT$ are 
enhancement of the power factor ($PF = \sigma S^2$), which can be accomplished by band-structure 
engineering\cite{pei2011convergence,pei2012band,liu2012convergence,dehkordi2015thermoelectric,parker2015benefits}, or
reduction of the lattice thermal conductivity through alloying and  
nanostructuring,\cite{hochbaum2008enhanced,boukai2008silicon,kanatzidis2009nanostructured,zhao2013high,vineis2010nanostructured} 
or by finding materials with 
intrinsically low $\kappa_\mathrm{latt}$.\cite{morelli2008intrinsically,he2016ultralow,gonzalez2018ultralow}
Though minimization of $\kappa_\mathrm{tot}$ is crucial, 
less attention has been paid to the carrier contribution to the thermal conductivity, $\kappa_\mathrm{el}$. 
The carrier concentration can be significant 
in doped TE materials with optimized carrier densities, so the 
effects of $\kappa_\mathrm{el}$ should not be ignored. However, 
minimization of $\kappa_\mathrm{el}$ through the reduction 
of $\sigma$ can be counter-productive due to the corresponding 
reduction in the power factor. 
In order to navigate the interdependence of the relevant properties, 
it is claimed that the reduction of $\kappa_\mathrm{el}$ can be best accomplished by 
minimizing the Lorenz number, $\Lambda = \kappa_\mathrm{el}/(\sigma T)$.\cite{mckinney2017search} 
Moreover, as pointed out by Mahan and Sofo\cite{mahan1996best},
$zT$ is always bounded by $\kappa^{0}/\kappa_\mathrm{latt}$, where $\kappa^{0}$ is 
the thermal conductivity due to carrier transport under isoelectrochemical conditions. 
Thus, the maximization of $\kappa^{0}$ allows for a higher upper limit for $zT$, 
an important result that has not been fully exploited given the difficulty in 
accurately calculating $\kappa^{0}$.
Despite the complexity arising from the interdependence of 
all the transport properties that contribute to $zT$, impressive 
progress has been made 
and new high-performance TE materials are 
continuously emerging.\cite{biswas2012high,liu2012copper,fu2016enhanced,olvera2017partial,cheng2017new,ma2020alpha,roychowdhury2021enhanced}

In the search for high $zT$ materials, bulk crystals 
with two-dimensional (2D) layered structures have attracted attention 
in recent years due to their high anisotropy and 
improved electrical conductivity along in-plane directions.\cite{terasaki1997large,rhyee2009peierls,ohta2018high,cheng2019optimal}
The recent discovery of a high $zT$ value for
intrinsic\cite{zhao2014ultralow} and p-doped SnSe\cite{zhao2015ultrahigh} are examples that have
boosted the interest in high-efficiency bulk TE materials.
Meanwhile, the out-of-plane direction had been ignored 
due to the generally low electrical conductivity along the stacking axis, 
even though it is accompanied by intrinsically low lattice thermal conductivity. 
This perspective changed recently when an outstanding 
TE performance with $zT = 2.8$ at 773~K was reported for n-doped SnSe in the out-of-plane direction .\cite{chang20183d}
The authors attributed the 
outstanding performance to two main factors: $i)$ the delocalization of Sn and Se p electrons 
close to the conduction band minimum that enables 
high conductance between Sn and Se atoms along the out-of-plane direction, and $ii)$ 
a continuous phase transition from \textit{Pnma} to \textit{Cmcm}, starting at 600~K and 
completing by $\sim$810~K, that results in the divergence of two nearly degenerate conduction bands, causing a 
decrease in the band mass and consequently higher conductivity.  
However, such an argument, on the basis of a two-band model, cannot be fully reconciled 
with the fact that the observed Seebeck coefficients do not decrease, 
as would be expected if the average band mass were to decrease.\cite{yang2015outstanding,kutorasinski2015electronic} 

Optimizing the carrier density by chemical doping is 
one of the most important strategies for improving TE properties of semiconductors.
For SnSe, hole-doping by p-type dopants such as 
Ag~\cite{chen2014thermoelectric,leng2016optimization,peng2016broad} and Na~\cite{peng2016broad} has 
led to increased values of $zT$ compared to the undoped material over a broad range in temperature.
Likewise, electron-doping by n-type dopant atoms, 
such as I\cite{zhang2015studies}, Bi\cite{duong2016achieving}, 
and Br\cite{chang20183d}, has also led to the enhancement of $zT$, with the 
latter yielding an impressive $zT= 2.8$ for the out-of-plane direction in \textit{Pnma}-SnSe at 773~K. 
Unraveling
the microscopic origin of the outstanding TE performance 
of n-doped SnSe will be extremely helpful in advancing 
the search for improved TE materials.  
To this end, we have conducted an extensive first-principles 
investigation of the electron-phonon (e-p) coupling 
and related properties, which were combined with calculations 
based on a semi-empirical theory for ionized impurity scattering,\cite{brooks1955theory,chattopadhyay1981electron} in order to 
calculate TE transport properties in the out-of-layer direction of n-doped \textit{Pnma}-SnSe within 
the Boltzmann transport equation (BTE) framework. For comparison, we 
calculated the same properties for the out-of-layer direction of p-doped \textit{Pnma}-SnSe 
with similar carrier density.

Our first-principles calculations of the e-p coupling are based 
on the dual interpolation technique\cite{chaves2020boosting} 
for computing e-p matrix elements using density functional 
theory (DFT)\cite{hohenberg1964inhomogeneous,kohn1965self} band structures and
density functional perturbation theory (DFPT)\cite{baroni2001phonons}
phonon dispersions.
We determined the dominant scattering mechanisms as 
functions of carrier energy and temperature, as 
well as the average electronic group velocities, which allow us to 
predict the overall transport properties and understand 
the origin of the high $zT$ value as well as the
enhancement of $zT$ above 600~K for n-doped SnSe. 
In particular, the total relaxation time, $\tau_\mathrm{tot}$, increases with temperature for n-doping, 
a feature that maintains a high PF at temperatures above 600~K, 
while simultaneously reducing $\kappa_\mathrm{el}$ even faster, leading an ultrahigh-$zT = 3.1$ at 807~K. 
Additionally, given the accuracy of our calculations, we explain the roles played by $\kappa^{0}$ 
and $\kappa_\mathrm{el}$ in the determination of $zT$. 
Our results show that $\kappa^{0}/\kappa_\mathrm{latt}$ indeed represents the upper limit
for $zT$, whereas the difference between calculated $zT$ and $\kappa^{0}/\kappa_\mathrm{latt}$
is directly proportional to $\kappa_\mathrm{el}$.

\section{Theoretical Approach}

We have performed the most comprehensive first-principles calculations 
of thermoelectric properties to date. In particular, as described below, 
we combine non-polar scattering of carriers by acoustic and optical phonons, 
polar scattering within the Fr{\"o}hlich theory 
including Ehrenreich screening, and the scattering by ionized impurities including 
non-parabolic contributions. 

Starting from the semiclassical BTE within the
relaxation time approximation (RTA),~\cite{madsen2006boltztrap,chaves2021investigating}
the key quantity required to calculate thermoelectric transport 
properties is the momentum- and band-resolved transport distribution kernel,
$\Sigma_{\alpha,\beta} (n,{\bf{k}}) = e^2\tau_{n,{\bf{k}}} v_{\alpha}(n,{\bf{k}}) v_{\beta} (n,{\bf{k}})$, 
where $e$ is the absolute electric charge, $\tau_{n,{\bf{k}}}$ is the total relaxation time and $v_\alpha(n,{\bf{k}})$ is the $\alpha$-component of the
average group velocity for a given electronic state with band index $n$ and wave vector ${\bf{k}}$. 
The energy projected transport function can then be defined over an energy grid with spacing $d\epsilon$ as 
\begin{equation}
\label{kernel}
\Sigma_{\alpha,\beta} (\epsilon) = \frac{1}{N_{{\bf{k}}}} \sum{\Sigma_{\alpha,\beta} (n,{\bf{k}})}\frac{\delta(\epsilon - \epsilon_{n,{\bf{k}}})}{d\epsilon}
\end{equation}
where $N_{{\bf{k}}}$ is the number of {\bf{k}} points sampled and $\epsilon_{n,{\bf{k}}}$ is the band energy. 
The temperature ($T$) and chemical potential ($\mu$) dependent transport tensors can then be calculated 
as an energy integral of the different energy moments:
\begin{equation}
\label{1}
\sigma_{\alpha,\beta} (T,\mu) = \frac{1}{\Omega} \int{\Sigma_{\alpha,\beta} (\epsilon) \left(-\frac{\partial f_{\mu}(T,\epsilon)}{\partial \epsilon} \right)} d\epsilon~,
\end{equation}

\begin{equation}
\label{2}
\phi_{\alpha,\beta} (T,\mu) = \frac{1}{eT\Omega} \int{\Sigma_{\alpha,\beta} (\epsilon)\left(-\frac{\partial f_{\mu}(T,\epsilon)}{\partial \epsilon} \right) (\epsilon - \mu)} d\epsilon~,
\end{equation}

\begin{equation}
\label{3}
\kappa^0_{\alpha,\beta} (T,\mu) = \frac{1}{e^2T\Omega} \int{\Sigma_{\alpha,\beta} (\epsilon)  \left(-\frac{\partial f_{\mu}(T,\epsilon)}{\partial \epsilon} \right)(\epsilon - \mu)^2} d\epsilon~,
\end{equation}
where $\Omega$ is the volume of the unit cell and $f_{\mu}(T,\epsilon)$ is 
the Fermi-Dirac distribution function. 
The thermoelectric transport coefficients for each crystallographic direction can then be derived 
from the above tensors by taking the trace as performed in reference~\citenum{chaves2021investigating}, 
where $\sigma \equiv \sigma_{\alpha,\beta} (T,\mu)$ is the electrical conductivity, 
$S \equiv S_{\alpha,\beta} (T,\mu) = \phi_{\gamma,\alpha}(\sigma^{-1}_{\gamma,\beta})$ is the Seebeck coefficient and 
$\kappa_\mathrm{el} \equiv \kappa_{\alpha,\beta}^\mathrm{el} (T,\mu) = \kappa^0_{\alpha,\beta} - T\phi_{\alpha,\gamma}(\sigma^{-1}_{\delta,\gamma})\phi_{\delta,\beta}$ is 
the thermal conductivity due to carrier transport at zero electric current, 
calculated from $\kappa^0$, which is the thermal conductivity 
due to the carrier transport under isoelectrochemical conditions. 

Since SnSe is a polar semiconductor, carriers are expected 
to be predominantly scattered via 
interactions with phonons at finite temperature and ionized impurities, especially 
in the case of doped SnSe.   
In our calculations, the relaxation time (RT) for the e-p scattering
is related to the imaginary part of the momentum- and band-resolved Fan-Migdal electron
self-energy\cite{giustino2017electron,ponce2018towards,ponce2020first}, 
\begin{equation}
  \frac{1}{\tau_{n,{\bf{k}}}} = 2 \mathrm{Im\,} \Theta_{n,{\bf{k}}}(\epsilon=0,T),
 \end{equation}
with\noindent
\begin{flalign} 
\label{imag}
 \begin{split}
& \mathrm{Im\,} \Theta_{n,{\bf{k}}}(\epsilon,T) = \pi \sum_{m,\theta} \int_{BZ} \frac{d{\bf{q}}}{\Omega_{BZ}} |g_{mn,\theta}({\bf{k},{\bf{q}}})|^2 \\
&       \times \Bigg[\left[n_{{\bf{q}}\theta}(T) + f_{m{\bf{k}}+{\bf{q}}}(T)\right]\delta(\epsilon - (\epsilon_{m{\bf{k}}+{\bf{q}}} - \epsilon_F) + \omega_{{\bf{q}}\theta}) \\
&        + [n_{{\bf{q}}\theta}(T) + 1 - f_{m{\bf{k}}+{\bf{q}}}(T)]\delta(\epsilon - (\epsilon_{m{\bf{k}}+{\bf{q}}} - \epsilon_F) - \omega_{{\bf{q}}\theta})\Bigg]~,
\end{split}
\end{flalign}
where $\epsilon_F$ is the Fermi energy calculated using DFT at 0~K, 
$g_{mn,\theta}({\bf{k},{\bf{q}}}) = \braket{\Psi_{m{\bf{k}}+{\bf{q}}}|\partial_{{\bf{q}}\theta} V_{KS}(r)|\Psi_{n{\bf{k}}}}$ 
are the e-p coupling matrix elements calculated within DFPT, 
$\ket{\Psi_{n{\bf{k}}}}$ are Kohn-Sham (KS) orbitals 
and $\partial_{{\bf{q}}\theta} V_{KS}$ 
corresponds to the change of the KS potential upon a
phonon perturbation with momentum {\bf{q}} and branch index $\theta$,  
$n_{{\bf{q}}\theta}(T)$ is the Bose-Einstein distribution function,
$\Omega_{BZ}$ is the Brillouin zone (BZ) volume, $m$ and $n$ are 
band indices, and 
$\omega_{{\bf{q}}\theta}$ are phonon eigenfrequencies.

Scattering of charge carriers by the electric polarization 
caused by longitudinal optical (LO) phonons can be prominent in polar materials such as SnSe. 
This polar mode scattering was first discussed
by Fr{\"o}hlich~\cite{frohlich1937h} and Callen~\cite{callen1949electric},
while Howarth and Sondheimer~\cite{howarth1953theory} developed the theory of polar mode scattering
by treating electrons as charge carriers on a simple parabolic
conduction band. First-principles treatment of the Fr{\"o}hlich interaction is not amenable
to Wannier-Fourier (W-F) interpolation, since the long 
range Fr{\"o}hlich interaction~\cite{rohlfing2000electron}
requires a very large number of e-p matrix elements to attain convergence. 
We account for the polar mode scattering following the
first-principles method of Verdi and Giustino~\cite{verdi2015frohlich}, in which the polar
singularity is treated by separating the e-p matrix elements into short- and long-range parts: 
$g_{mn,\theta}({\bf{k},{\bf{q}}}) = g^{{\mathcal{S}}}_{mn,\theta}({\bf{k},{\bf{q}}}) + g^{{\mathcal{L}}}_{mn,\theta}({\bf{k},{\bf{q}}})$.
The short-range part is well behaved within W-F interpolation while 
the long-range part can be treated by using an analytical formula based on the
Vogl model~\cite{pellegrini2016physics,bostedt2016linac,vogl1976microscopic}
\begin{multline}
\label{long}
g_{mn,\theta}^{\mathcal{L}}({\bf{k}},{\bf{q}}) = i \frac{e^2}{\Omega\epsilon_0} \sum_{\kappa}
 \left(\frac{\hslash}{2 N{M_{\kappa} \omega_{{\bf {q}}\theta}}}\right)^{\!\!\frac{1}{2}} \\
 \times \sum_{{\bf G}\ne -{\bf q}} 
 \frac{ ({\bf {q}}+{\bf {G}})\cdot{\bf {Z}}^*_\kappa \cdot {\bf {e}}_{\kappa\theta}({\bf {q}}) } 
 {({\bf {q}}+{\bf {G}})\cdot\bm\zeta_\infty\!\cdot({\bf {q}}+{\bf {G}})} \\
 \times \braket{\Psi_{m{\bf{k}}+{\bf{q}}}|e^{i({\bf{k}}+{\bf{q}})\cdot{\bf{r}}}|\Psi_{n{\bf{k}}}}~, 
\end{multline}
in which $M_{\kappa}$ corresponds to the mass of atom $\kappa$,  
$N$ is the number of unit cells in the Born-von K{\'a}rm{\'a}n supercell, 
${\bf{G}}$ is a reciprocal lattice vector, ${\bf{Z}}^*=Z^*_{\alpha,\beta}$ 
is the Born effective charge tensor, ${\bf {e}}_{\kappa\theta}({\bf {q}})$ 
is a phonon eigenmode normalized within the unit cell, $\bm\zeta_\infty = \zeta^{\infty}_{\alpha,\beta}$ 
corresponds to the high-frequency dielectric constant tensor, $\epsilon_0$ is the vacuum permittivity, and $\hslash$
is the reduced Planck constant.  
$\braket{\Psi_{m{\bf{k}}+{\bf{q}}}|e^{i({\bf{k}}+{\bf{q}})\cdot{\bf{r}}}|\Psi_{n{\bf{k}}}} = \left[ U_{{\bf {k}}+{\bf {q}}}\:U_{{\bf {k}}}^{\dagger} \right]_{mn}$ are phase factors given in terms of rotation matrices, $U_{{\bf {k}}+{\bf {q}}}$, 
that appear in the definition of the maximally localized Wannier functions (MLWFs).\cite{marzari1997maximally} 

The above expression represents a first-principles generalization of the 
Fr{\"o}hlich coupling within the theory of polarons\cite{frohlich1954electrons} and 
analogously treats the problem of a single electron added to a polar insulator, 
without considering the screening of the e-p coupling caused by a finite carrier density.
The generalization to include
screening effects beyond Fr{\"o}hlich theory 
was developed by Ehrenreich~\cite{ehrenreich1959screening}. 
In the quasi-static approximation, free carriers that are present in the sample screen out
the electric field produced by optical vibrations, resulting in both a weakening of the
e-p coupling as more carriers are added to the system, and a
shift of the frequency of the
longitudinal optical mode~\cite{ehrenreich1959screening}. 
The former effect weakens the e-p matrix element by a factor of $1-(r_{\infty}{\bf{q}})^{-2}$, 
where $r_{\infty}$ is the screening radius given by 
\begin{equation}
\label{r0inf}
r{_{\infty}}^{-2}(n,{\bf{k}}) = \frac{4\pi{e^2}}{\zeta{_{\infty}}}\int{\left(-\frac{\partial f_{\mu}(T,\epsilon)}{\partial \epsilon_{n,{\bf{k}}}}\right) g(\epsilon) d\epsilon}~,
\end{equation}
and $g(\epsilon)$ is the density of states (DOS), given by
\begin{equation}
\label{dos2}
g(\epsilon) = \int \sum_n \delta(\epsilon - \epsilon_{n,{\bf{k}}}) \frac{d{\bf{k}}}{8\pi^3} = \frac{1}{\Omega N_{{\bf{k}}}}\sum_{n,{\bf{k}}} \frac{\delta(\epsilon - \epsilon_{n,{\bf{k}}})}{d\epsilon}~.
\end{equation}

The latter effect of the screening leads to an eigenfrequency shift of the
LO phonons given by
\begin{equation}
\label{freq_shift}
{(\omega^{LO}})^2 = {(\omega^{TO}})^2\left(\frac{\zeta_0/\zeta_{\infty}+ (r_{\infty}{\bf{q}})^{-2}}{1+(r_{\infty}{\bf{q}})^{-2}}\right)~,
\end{equation}
where $\omega^{TO}$ is the transverse optical (TO) mode eigenfrequency.
The eigenfrequency of the LO vibration 
is strongly reduced, altering the e-p matrix elements~\cite{ravich1971scattering}.
Therefore, Ehrenreich quasi-static screening modifies  the polar scattering RT by the following band-dependent factor:
\begin{multline}
\label{screen_pol}
 \begin{split}
& F_\mathrm{pol}(n,{\bf{k}}) = \left[1 -\frac{1}{2(r_{\infty}(n,{\bf{k}})\cdot{\bf{k}})^2}\right.
\\
& \left.\times \ln[1+4(r_{\infty}(n,{\bf{k}})\cdot{\bf{k}})^2] + \frac{1}{1+4(r_{\infty}(n,{\bf{k}})\cdot{\bf{k}})^2}\right]^{-1}~.
\end{split} 
\end{multline}
Combining Eqs.~\ref{imag}, \ref{long}, and \ref{screen_pol}, we arrive at expressions for the RT corresponding to both 
non-polar ($\tau_\mathrm{npol}$) and screened polar ($\tau_\mathrm{pol}$) phonon scattering.  
The non-polar e-p RT is given by 
\begin{equation}
\label{npol}
\frac{1}{\tau_\mathrm{npol}(n,{\bf{k}})} = 2\, \mathrm{Im\,} \Theta_{n,{\bf{k}}}[\epsilon = 0,T,g^{{\mathcal{S}}}_{mn,\theta}({\bf{k},{\bf{q}}})]~,
\end{equation} 
and the screened polar e-p RT is given by 
\begin{equation}
\label{pol}
\frac{1}{\tau_\mathrm{pol}(n,{\bf{k}})} = 2\, \mathrm{Im\,} \Theta_{n,{\bf{k}}}[\epsilon = 0,T,g^{{\mathcal{L}}}_{mn,\theta}({\bf{k},{\bf{q}}})]\times F_\mathrm{pol}(n,{\bf{k}})~.
\end{equation}   

The scattering by ionized impurities
is calculated on the basis of the theory developed by 
Brooks and Herring.~\cite{brooks1955theory,chattopadhyay1981electron}
This framework 
neglects the effects of the impurities on the electron energy 
levels and wave functions and assumes that carriers are scattered independently
by dilute concentrations of ionized centers randomly distributed within the material.
It constitutes an accurate yet simple description, neglecting complex effects 
such as coherent scattering from pairs
of impurity centers, which requires a quantum transport theory~\cite{moore1967quantum}.
\begin{figure*}
        \centering
        \includegraphics[width=1\textwidth]{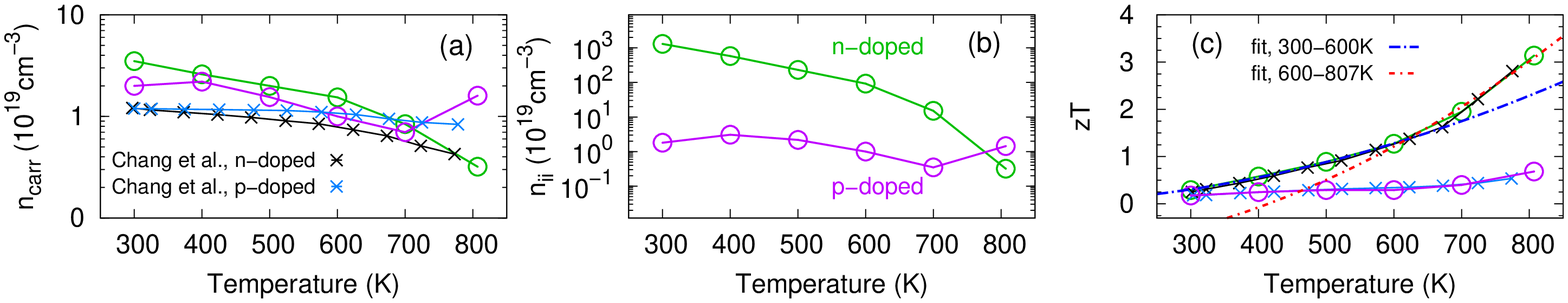}
        \caption{$a)$ Calculated carrier density ($n_\mathrm{carr}$) in comparison to experimental values inferred from Hall measurements, as reported by Chang et al.,\cite{chang20183d}
                 $b)$ calculated ionized impurity concentration ($n_\mathrm{ii}$),
                 and $c)$ the resulting figure of merit ($zT$) for p- and n-doped SnSe, along with experimental results reported by 
                 Chang et al.\cite{chang20183d}. The quadratic fits to low temperature points only (300-600~K, blue dot-dashed line) and high temperature points only (600-807~K, red dot-dashed line) 
                 highlight the enhancement of $zT$ above 600~K for n-doped SnSe.} 
        \label{n_carr}
\end{figure*}
Following Refs.~\citenum{chaves2021investigating} and \citenum{askerov2009thermodynamics}, the carriers are assumed to scatter off a screened Coulomb potential and the Born
approximation is used to evaluate transition probabilities. 
Accordingly, the RT for impurity scattering is given by 
\begin{equation}
\label{tau_imp}
\tau_\mathrm{imp}(n,{\bf{k}}) = \frac{\hslash\zeta{_0}{^2}}{2{\pi}{e^4}{n_\mathrm{ii}}F_\mathrm{imp}(n,{\bf{k}})}{\bf{k}}^2 \left\vert\frac{\partial \epsilon_{n,{\bf{k}}}}{\partial {\bf{k}}}\right\vert~, 
\end{equation}
where $n_\mathrm{ii}$ is the concentration of ionized impurities and 
\begin{equation}
F_\mathrm{imp}(n,{\bf{k}}) = \ln(1+\eta) - \frac{\eta}{1+\eta} 
\end{equation}
is the screening function, with $\eta = (2{\bf{k}}\cdot r_0(n,{\bf{k}}))^2$. Here 
$r_0$ is the static screening radius given by Eq.~\ref{r0inf}, 
but now screened by the static dielectric constant, $\zeta_0$.

The average electronic group velocities are calculated from a Fourier interpolation of the
band structure expanded in terms of star functions\cite{shankland1971interpolation,koelling1986interpolation,chaves2020boosting}
\begin{multline}
\label{velocities2}
\frac{\partial \epsilon_{n,{\bf{k}}}}{\partial {\bf{k}}} \equiv v({n,{\bf{k}}}) 
\approx \frac{i}{n_s}\sum_{m=1}^M a_m \sum_{\{\upsilon\}}(\upsilon {\bf{R}}_{m}) \exp [i(\upsilon {\bf{R}}_{m})\cdot{\bf{k}}]~, 
\end{multline}
with the sum running over all $n_s$ point group symmetry operations $\{\upsilon\}$
on the direct lattice translations, ${\bf{R}}_m$. $M$ is the number of star functions per ${\bf{k}}$ point and 
the $a_m$ are the Fourier coefficients
of the expansion of the band structure in terms of star functions.\cite{chaves2020boosting}

Once the RT for each of the three scattering process is computed,
$\tau_\mathrm{tot}(n,{\bf{k}},\mu,T)$, the total RT that enters in the TE transport calculations,
can be determined from Mathiessen's rule:
\begin{equation}
\label{Mathiessen}
\frac{1}{\tau_\mathrm{tot}} = \frac{1}{\tau_\mathrm{npol}}
+ \frac{1}{\tau_\mathrm{pol}} + \frac{1}{\tau_\mathrm{imp}}~.
\end{equation}
This is justified if the scattering mechanisms are approximately independent. 
The temperature dependence of the RT is given indirectly through the phonon and electron distributions 
within Eq.~\ref{imag}. Additionally, for $\tau_\mathrm{pol}$ and $\tau_\mathrm{imp}$, $T$ and $\mu$ dependence enters 
implicitly through their respective screening radii ($r_{\infty}$ and $r_0$) as defined in Eq.\ref{r0inf}. This dependence on $\mu$ allows for the study of
doped materials, which are important for the optimization of $zT$ for thermoelectric applications.  

\section{Computational details}

At room temperature, SnSe crystallizes in a layered orthorhombic structure
with the \textit{Pnma} space group and 8 atoms in the unit cell.
We used the Quantum Espresso package\cite{giannozzi2009quantum,giannozzi2017advanced} along with fully relativistic, optimized, 
norm-conserving Vanderbilt pseudopotentials\cite{hamann2013optimized,van2018pseudodojo} to calculate the electronic structure using DFT and determine the vibrational and e-p matrix elements within the DFPT framework. 
We used the generalized gradient approximation (GGA)
for the exchange-correlation functional within the formulation of Perdew-Burke-Ernzerhof (PBE)\cite{perdew1996generalized}.
Monkhorst-Pack grids of 20$\times$20$\times$10 for ${\bf{k}}$-point sampling and
a kinetic energy cutoff of 90 Ry were employed to ensure the
convergence of total energy in DFT calculations.
As expected, DFT-GGA calculations underestimate the band gap, so in order to compare with experimental data we applied the commonly used
scissor operator\cite{chen2016understanding,li2019resolving} to rigidly shift the conduction bands
in order to match the experimental value of 0.86~eV for the SnSe band gap.~\cite{zhao2014ultralow} 
The interlayer interactions in \textit{Pnma}-SnSe arise from weak van der Waals forces between Se and Sn atoms separated by $\sim$3.50 \AA{} in the out-of-plane direction.
In order to capture such weak interactions between the layers, we added van der Waals (vdW) corrections according to the D3 approach as proposed by Grimme \textit{et al.}~\cite{grimme2010consistent}.
We used the experimental structure~\cite{adouby1998structure} as the starting configuration and
relaxed the lattice parameters and atomic positions until
all atomic force components were smaller than 1~meV/\AA{}, yielding the following lattice parameters: $a = 11.79\:$\AA, $b = 4.52$ \AA{} and $c = 4.22$ \AA{}. These $T=0$ theoretical results differ by only $1.28$\%, $3.10$\% and $0.29$\% from the experimental lattice
parameters measured at $T=673$~K.\cite{adouby1998structure}
\begin{figure*}
        \centering
        \includegraphics[width=1\textwidth]{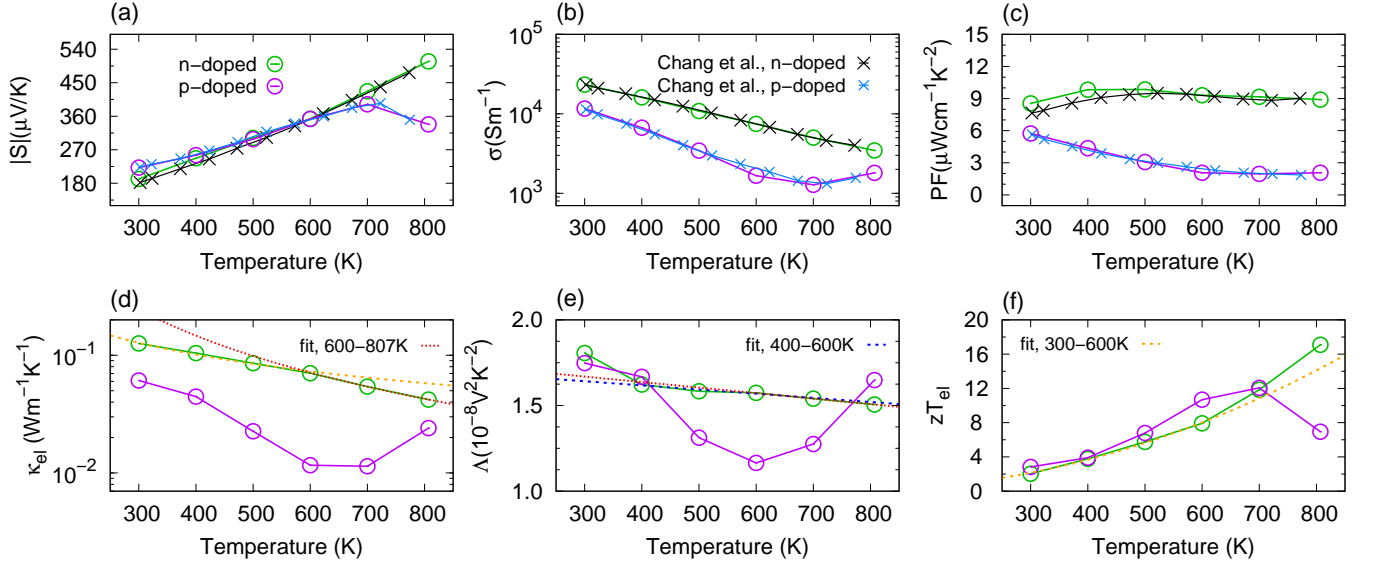}
        \caption{Calculated thermoelectric properties of p- and n-doped SnSe for $T =$ 300-807~K
                 in comparison with available experimental data reported by Chang et al.\cite{chang20183d} up to 773~K.
                 $a)$ Seebeck coefficient (S), $b)$ electrical conductivity ($\sigma$), $c)$ power factor ($PF$), $d)$ thermal conductivity
                 due to the carrier transport at zero electric current (${\kappa}_\mathrm{el}$), $e)$ Lorenz function ($\Lambda$) 
                 and $f)$ electronic figure of merit
                 ($zT_\mathrm{el}$). The fits to points in different temperature ranges (orange for 300-600~K, blue for 400-807~K and red for 600-807~K) highlight the enhancement  
                 (faster decrease) of $S$ and $zT_\mathrm{el}$ (${\kappa}_\mathrm{el}$ and $\Lambda$) above 600~K for n-doped SnSe.}
        \label{therm_prop}
\end{figure*}

The RTs arising from e-p scattering, including both contributions of 
non-polar scattering and screened polar scattering as presented in the previous section, 
were calculated using our in-house Turbo-EPW implementation,\cite{chaves2020boosting} 
which utilizes the dual interpolation technique 
based on MLWFs and symmetry-adapted star functions for efficient interpolation of e-p 
scattering matrix elements onto very fine meshes of electron (${\bf{k}}$) and phonon (${\bf{q}}$) wave vectors.
In the present case this interpolation allowed for calculations based on $\sim$3 billion ${\bf{k/q}}$ pairs. 
The first W-F interpolation, using MLWFs determined 
by the Wannier90 code,\cite{mostofi2008wannier90} leads to a phonon grid of 40$\times$40$\times$20 ${\bf{q}}$ points 
starting from a coarse grid of 4$\times$4$\times$2 points. Subsequently, starting from an initial coarse grid of 12$\times$12$\times$6 $\mathbf{k}$ points, $M=10$ star functions were used 
for the second interpolation, resulting in a dense grid of 64$\times$60$\times$24 ${\bf{k}}$ points.   

As described in the previous section, we account for ionized impurity scattering by starting from the semiempirical model of Brooks and Herring and then using Fourier interpolation of the DFT band structure in order to avoid the approximation of parabolic bands.\cite{chaves2021investigating}
We used the experimentally determined values\cite{chandrasekhar1977infrared} for the static 
and high-frequency dielectric constants, $\zeta_0 = 45$ and $\zeta_{\infty} = 13$, respectively.  
The same value of $M=10$ star functions used in the e-p calculations was employed in the computation 
of $\tau_\mathrm{imp}$ in order to have a consistent grid for integration. 
Finally, Mathiessen's rule (Eq.\ref{Mathiessen}) yields $\tau_\mathrm{tot}$, which 
was used to compute TE transport coefficients based on a modified 
version of the BoltzTraP code.\cite{madsen2006boltztrap,chaves2021investigating}

\section{Results}

The comprehensive theoretical framework and accurate computational
implementation presented in the preceding sections allow us to untangle the 
microscopic factors behind the high zT value exhibited by out-of-layer n-doped \textit{Pnma}-SnSe. 

\subsection{Carrier density and ionized impurity concentration}

Following the iterative procedure described in 
reference\cite{chaves2021investigating}, we determined the carrier density ($n_\mathrm{carr}$) and ionized impurity concentration ($n_\mathrm{ii}$) at each temperature $T$ by requiring that the calculated Seebeck coefficients 
and electrical conductivities matched the measured values reported in the work of Chang \textit{et al.}\cite{chang20183d} for temperatures up to 773 K. For higher temperatures, $n_\mathrm{carr}$ and $n_\mathrm{ii}$ were determined based on a smooth extrapolation of $S$ and $\sigma$ with increasing $T$.
As shown in Fig.~\ref{n_carr}, the resulting $n_\mathrm{carr}$ is in good
agreement with the experimental values inferred from 
Hall measurements.\cite{chang20183d} It is important to note that 
Hall concentrations are determined assuming a single 
parabolic band and a Hall scattering factor of unity, which means they should not be considered 
to be the exact carrier densities. However, such measurements provide a reasonable 
estimate of $n_\mathrm{carr}$ and serve as a qualitative check of our 
determinations of $n_\mathrm{carr}$ and $n_\mathrm{ii}$. Experimental and 
calculated values of $S$, $\sigma$, and power factor, $PF=\sigma S^2$, are shown in Fig.~\ref{therm_prop}. 
The close match between calculated and measured values demonstrates that 
our theoretical framework is robust and can accurately describe the temperature 
dependence of these TE properties for reasonable values of $n_\mathrm{carr}$ and $n_\mathrm{ii}$.
\begin{figure*}[ht]
\centering
\includegraphics[width=0.95\textwidth]{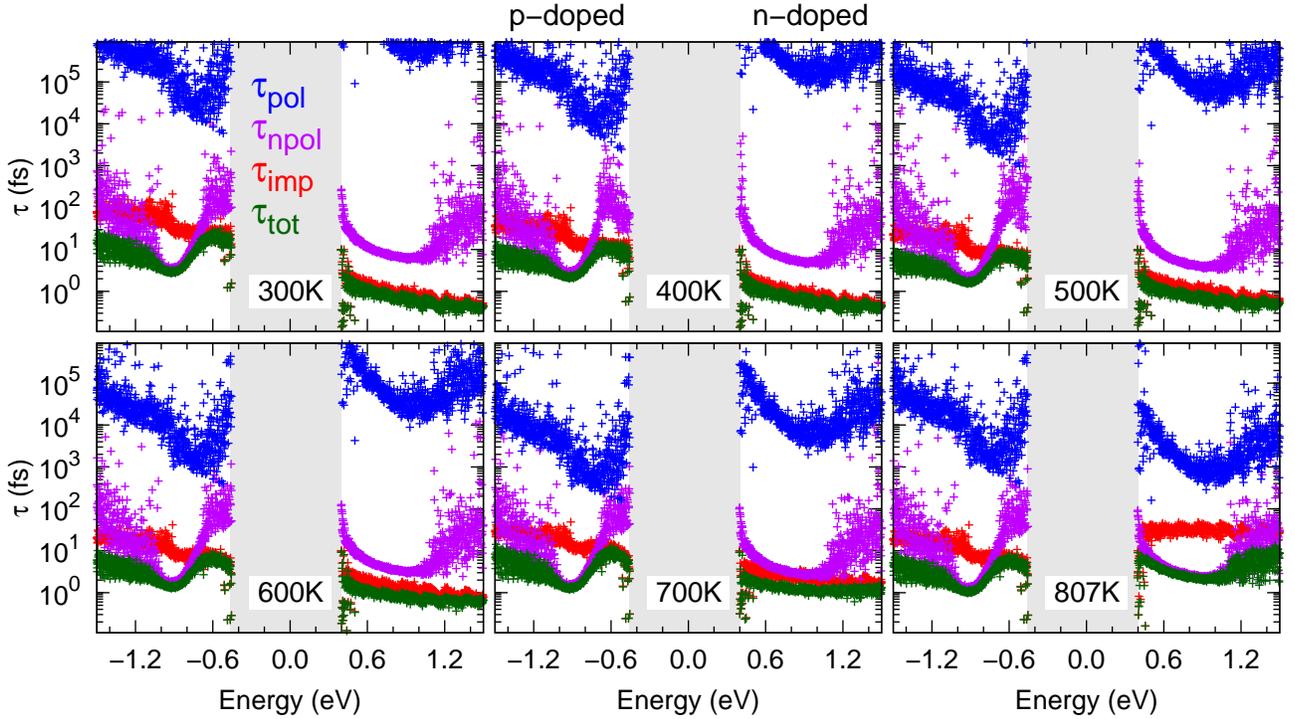}
\caption{Temperature dependence (300-807~K) of relaxation times (RTs) as a function of carrier energies
calculated for each scattering process in the out-of-layer direction of p- and n-doped SnSe:
non-polar scattering of acoustic and optical phonons ($\tau_\mathrm{npol}$, magenta),
screened polar scattering of optical phonons ($\tau_\mathrm{pol}$, blue), scattering by ionized impurities ($\tau_\mathrm{imp}$, red), and
the total RT ($\tau_\mathrm{tot}$, green) calculated using Mathiessen's rule.
The light grey rectangles represent the gap region with an energy gap of 0.86~eV.
}
\label{Tau}
\end{figure*}

In the temperature range of 400-700~K, the calculated $n_\mathrm{carr}$ 
and $n_\mathrm{ii}$ for both n- and p-doped SnSe gradually decrease with temperature.
This decrease correlates well 
with the measured Hall carrier concentrations.
For higher temperatures the carrier and ionized impurity concentrations in n-doped 
SnSe continue to decrease, whereas they increase for p-doped SnSe. 
This increase in carrier and ionized impurity concentration in p-doped SnSe can be attributed to 
the exponential intrinsic temperature-driven formation of 
Sn vacancies\cite{dewandre2016two,huang2017first,zhou2018influence,chaves2021investigating}, $V_{Sn}^{-2}$.
At 600~K this temperature-driven vacancy formation starts to exceed the concentration of
vacancies formed during growth  
and add additional holes  
beyond those generated by the extrinsic defects due to p-doping. 

For the n-doped case the additional holes from Sn vacancies 
would lead to a further reduction in the carrier density, 
as observed in Fig.~\ref{n_carr}(a), since it is expected 
that these vacancies will capture electrons. 
In the work of Chang \textit{et al.}\cite{chang20183d}, the n-type dopant impurity is Br, 
which substitutes for Se atoms and forms $Br_{Se}^+$ charged defects.
Since the Fermi level is close to the bottom of the conduction band, 
the formation energy of $V_{Sn}^{-2}$ is low because they capture electrons. 
Even the undoped material is a p-type semiconductor due to the $V_{Sn}^{-2}$ 
that are formed during growth and remain trapped 
in the structure as the material cools down\cite{zhou2018influence}. 
They introduce defect levels close to the valence band maximum 
and thus provide partial compensation in the n-doped material.
As temperature goes up the concentration of $V_{Sn}^{-2}$ goes up as well. Then, it is expected 
that the whole concentration of ionized impurities would go up. 
However, our calculations indicate that the concentration
of ionized impurities actually decreases with increasing temperature (Fig.~\ref{n_carr}(b)), 
which can be explained by the likely formation of $(2Br_{Se}^+)$-$V_{Sn}^{-2}$ complexes, 
which are neutral and electrically inert. The scattering of carriers              
by these neutral impurities is not relevant because their concentration remains vanishingly small 
when compared to the concentration of ionized impurities.\cite{brooks1955theory,canali1975electron,li2012semiconductor} 

\subsection{Calculated Thermoelectric Properties}\label{therm}

The magnitudes of the Seebeck coefficient for p- and n-doped SnSe are nearly identical
up to $\sim$700~K, but then diverge for higher temperatures.
While the Seebeck coefficient continues increasing for n-doped SnSe,
for p-doped SnSe it decreases for higher temperatures.
The decrease in $S$ for the p-doped case is a consequence of two main factors: 
$i)$ the increase of $n_\mathrm{carr}$, 
likely caused by $V_{Sn}^{-2}$ formation and associated generation of holes; and
$ii)$ the increase of hole conduction, despite the associated increase 
in $n_\mathrm{ii}$ that leads to increased scattering by ionized impurities.
The onset of temperature-driven formation of $V_{Sn}^{-2}$  
also impacts n-doped SnSe by capturing electrons in the system. 
This leads $n_{carr}$ to decrease faster causing an enhancement in $S$ beyond $\sim600$~K. 

The electrical conductivity exponentially decreases with temperature for 
both p- and n-doped SnSe up to 700~K, and continues decreasing for the n-doped material, 
whereas the conductivity abruptly increases for the p-doped material. 
This increase is a direct consequence of the rise in $n_\mathrm{carr}$, 
despite the decrease in the relaxation time.
One important observation is that $\sigma$ decays slower for the n-doped material 
due to the relaxation time behaviour. 
Though the high-temperature behaviour of $S$ and $\sigma$ are quite 
different for the n- and p-doped cases, they compensate in such a way that 
the $PF$ is quite flat over the whole temperature range 
above 600~K for both dopings, as shown in Fig.~\ref{therm_prop}(c). 
This is also a consequence of the relaxation time behaviour, which will be discussed below. 
Another important feature to observe is that the n-doped SnSe presents 
a higher $PF$ than the p-doped case.

Having used $S$ and $\sigma$ to determine $n_\mathrm{carr}$ 
and $n_\mathrm{ii}$, the thermal conductivity of the carriers at zero electric current 
($\kappa_\mathrm{el}$), Lorenz factor ($\Lambda = \kappa_\mathrm{el}/(\sigma T)$), and electronic 
figure of merit ($zT_{el} = S^2/\Lambda$) can be directly calculated 
without any additional input. The results are shown in Fig.\ref{therm_prop}(d)-(f).
For n-doped SnSe, $\kappa_\mathrm{el}$ decreases monotonically 
with temperature, which gives rise to a similar monotonic decrease in 
$\Lambda$ and increase in $zT_\mathrm{el}$. On the other hand, the behaviour 
of $\kappa_\mathrm{el}$ for the p-doped material is decidedly non-monotonic, 
reaching a minimum between 600 and 700~K, which leads to a minimum in $\Lambda$ 
and maximum in $zT_\mathrm{el}$ near the same temperature. 

\begin{figure*}[ht]
\centering
\includegraphics[width=0.95\textwidth]{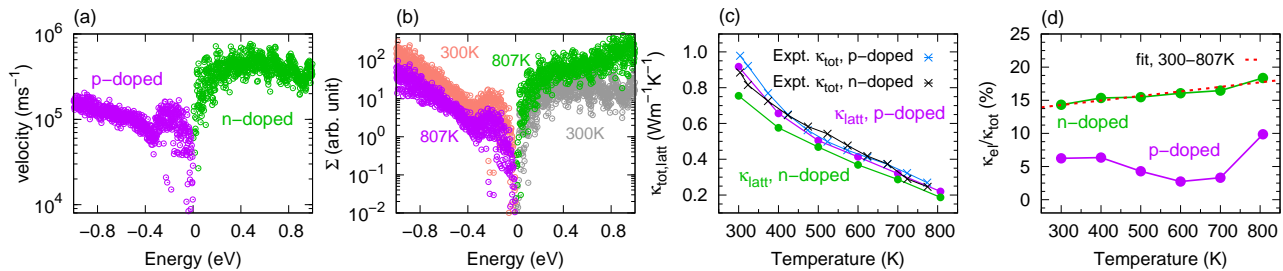}
\caption{$a)$ Electronic velocity $v(\epsilon)$ and $b)$ the transport distribution function, 
         $\Sigma(\epsilon)$ at 300 and 807~K, as a function 
         of the carrier energies for n- and p-doped SnSe, where both quantities are plotted in 
         relation to their respective band edges placed at 0~eV. $c)$ Lattice ($\kappa_\mathrm{latt}$) 
         and total ($\kappa_\mathrm{tot}$) thermal conductivities and $d)$ the contribution
         of carrier transport to the total thermal conductivity at zero electric current as a 
         function of temperature for n- and p-doped SnSe. 
         The red dashed linear fit (300-807~K) highlights the monotonic increase of $\kappa_\mathrm{el}$/$\kappa_\mathrm{tot}$.}
\label{veloc}
\end{figure*}

Additionally, we then determined $\kappa_\mathrm{latt}$ by 
subtracting the calculated $\kappa_\mathrm{el}$ from the experimental $\kappa_\mathrm{tot}$ 
as measured by Chang et al.\cite{chang20183d}
Figure~\ref{veloc} shows $\kappa_\mathrm{latt}$ and $\kappa_\mathrm{tot}$ 
for both p- and n-doped SnSe as a function of temperature. 
We observe that n-doped SnSe has a smaller $\kappa_\mathrm{latt}$, which can be attributed 
to its heavier doping, resulting in more ionized impurities which enhance phonon scattering. 
Since n-doped SnSe possesses a larger $\kappa_\mathrm{el}$, both systems have similar $\kappa_\mathrm{tot}$. 
By considering a $1/T$ extrapolation\cite{callaway1959model} of the calculated 
values of $\kappa_\mathrm{latt}$, 
we calculated $\kappa_\mathrm{tot}$ at 807~K, which was used to determine $zT$ for both systems (Fig.\ref{n_carr}(c)).
For the n-doped case, we obtained an ultrahigh-$zT = 3.1$ at 807~K, while for the p-doped case $zT = 0.7$. 
Since the total thermal conductivities for p- and n-doped SnSe are comparable, the leading cause 
behind the significantly higher $zT$ for the n-doped material is its higher $PF$, 
which is a consequence of its higher 
electrical conductivity. 
Importantly, above 600~K $zT$ starts growing faster with $T$. As we will explain below, this enhancement cannot be attributed
to the $PF$, which is almost constant in that range. 

\subsection{Dominant Scattering mechanisms}

We now consider the microscopic scattering mechanisms and their 
respective RTs, which determine the transport properties of p- and n-doped SnSe.
The RT contribution for each process, as well as for $\tau_\mathrm{tot}$, is defined 
as a function of the carrier energy by 
\begin{equation}
\label{Tau_eq}
\tau(\epsilon) = \frac{\sum_{n,{\bf{k}}} \tau_{n,{\bf{k}}} v_{n,{\bf{k}}} v_{n,{\bf{k}}} \delta (\epsilon-\epsilon_{n,{\bf{k}}})}{\sum_{n,{\bf{k}}} v_{n,{\bf{k}}} v_{n,{\bf{k}}} \delta (\epsilon-\epsilon_{n,{\bf{k}}})}~.
\end{equation}
Our results for $\tau(\epsilon)$ as a function of temperature 
for the out-of-plane direction of p- and n-doped SnSe 
are presented in Fig.\ref{Tau}. For both n- and p-doping the RT 
for polar e-p scattering is so large due to the strong screening 
that it will not have any effect 
on the TE properties, so we will not consider it further.

For the n-doped material, the dominant scattering over most of the 
temperature range is due to ionized impurities with a relaxation time, $\tau_\mathrm{imp}$, that 
smoothly decreases as a function of the carrier energy.  
The temperature dependence of the individual RTs is shown in Fig.~S1 in the Electronic Supplementary Information (ESI), 
where it is clear that $\tau_\mathrm{imp}$, and consequently $\tau_\mathrm{tot}$, 
both increase with temperature. 
This is due to the decrease of $n_\mathrm{ii}$, even though 
the screening of the impurities decreases as the charge density diminishes with temperature.  
On the other hand, since $\tau_\mathrm{npol}$ 
decreases with temperature, non-polar scattering 
by phonons plays a progressively larger role, finally becoming the dominant scattering mechanism at 807~K.

For the p-doped material, the dominant scattering mechanism 
depends on the carrier energy. The RT for ionized impurity scattering, $\tau_\mathrm{imp}$, 
steadily increases as the carrier energy moves away from the valence 
band maximum (VBM), dominating for both the lowest and highest energy carriers. 
In contrast, the RT for non-polar phonon scattering, $\tau_\mathrm{npol}$, 
has a pronounced "U" shape that dips below $\tau_\mathrm{imp}$ 
for intermediate carrier energies.
This has a strong influence on the high-$T$ TE properties, as will be discussed below.   
The temperature dependence of each RT is shown in Fig.~S2 of the ESI, 
where we observe that both $\tau_\mathrm{npol}$ and $\tau_\mathrm{imp}$, 
and thus also $\tau_\mathrm{tot}$, in general decrease with $T$. This behaviour of 
$\tau_\mathrm{imp}$ is opposite that of the n-doped case, and it is related to 
the evolution of $n_\mathrm{ii}$ with temperature.
Since $n_\mathrm{ii}$ is nearly constant with $T$ for the p-doped material, 
the temperature dependence of $\tau_\mathrm{imp}$ must be caused by the 
change in the screening radius. This feature is crucial for a proper description of the transport 
behaviour of p-doped SnSe, since $\tau_\mathrm{imp}$ dominates for energies close to VBM.

A direct comparison of the total RT for n-doped and p-doped SnSe is 
provided in Fig.~S3 in the ESI. At 300~K, $\tau_\mathrm{tot}$ for the 
p-doped material is about $\sim$10-20~fs at the VBM, almost one order of magnitude 
larger than that of n-doped SnSe, which has $\tau_\mathrm{tot}\sim$2~fs at the CBM. This difference
quickly decreases with increasing temperature, a direct consequence of 
the increase of $\tau_\mathrm{imp}$ for n-doped SnSe and simultaneous decrease of $\tau_\mathrm{npol}$ 
and $\tau_\mathrm{imp}$ for the p-doped material. In fact, 
at 807~K, the magnitudes of $\tau_\mathrm{tot}$ for both cases 
are within the same order of magnitude ($\sim$1-10fs). However, the 
dependence on carrier energy is quite different. While $\tau_\mathrm{tot}$ for 
the p-doped material exhibits a U-shaped behaviour, $\tau_\mathrm{tot}$ 
for n-doped SnSe decreases slowly and smoothly. 

The opposite $T$ dependence of $\tau_\mathrm{tot}$ for 
p- and n-doped doped SnSe is crucial for understanding  
their transport properties from a microscopic viewpoint. 
As $\tau_\mathrm{tot}$ increases with $T$ for the n-doped material, it slows down the decay rate of $\sigma$ as a function of $T$ and, 
since $\tau_\mathrm{tot}$ weakly influences $S$, it prevents the $PF$ from decreasing above 600~K, even though $n_\mathrm{carr}$ 
starts more rapidly decreasing due to the formation of Sn vacancies.  
Rather, the $PF$ stays roughly constant in the temperature range of 600-807~K, which
contributes to the excellent performance of n-doped SnSe, as will be discussed below. 
For the p-doped material, the opposite behaviour of $\tau_\mathrm{tot}$ greatly influences $\sigma$ by making it
decrease up to 700~K and then preventing it from increasing as quickly as the temperature rises to 807~K. Together with the behavior of $|S|$, this  
results in a $PF$ that is quite flat over the whole temperature range
above 600~K.   

\subsection{Average electronic group velocities}

Along with the RTs presented in the preceding section, the average electronic group 
velocity, $v(\epsilon)$, governs all TE properties through
the transport distribution function, $\Sigma(\epsilon)$, given by Eq.~\ref{kernel}.
$v(\epsilon)$ in the out-of-plane direction is given by 
\begin{equation}
 \label{vel}
 v(\epsilon) = \sqrt{\sum_{n,{\bf{k}}} |v_{n,{\bf{k}}}|^2 \delta (\epsilon-\epsilon_{n,{\bf{k}}})/\sum_{n,{\bf{k}}} \delta (\epsilon-\epsilon_{n,{\bf{k}}})}~.
 \end{equation}
Fig.~\ref{veloc}(a) shows $v(\epsilon)$ for p- and n-type SnSe, 
clearly demonstrating that the velocities for n-type SnSe are much higher than those for p-type SnSe. 
This characteristic is well known from the 
literature,\cite{ma2018intrinsic,li2019resolving,nassary2009electrical,yang2015outstanding,chang20183d,kutorasinski2015electronic} 
and is in line with observations that the electron effective masses in SnSe 
are much smaller than those of holes.\cite{nassary2009electrical,kutorasinski2015electronic} 
Yang et al.\cite{yang2015outstanding} pointed out that this difference is 
caused by two important factors. 
First, anti-bonding states formed by the interaction 
between s and p orbitals of Sn atoms with p orbitals of Se atoms push away the 
charge density of Sn atoms at the VBM, thus preventing hole transport.
Second, p orbitals of both Sn and Se atoms are much more 
delocalized at the CBM, enabling high conductance between Sn and Se 
atoms along the out-of-layer direction. Such delocalization has been 
further confirmed by DFT calculations and scanning tunneling microscopy, 
whose results indicate that the charge density 
tends to fill the out-of-plane interlayer region.\cite{chang20183d}

\begin{figure*}
        \centering
        \includegraphics[width=0.7\textwidth]{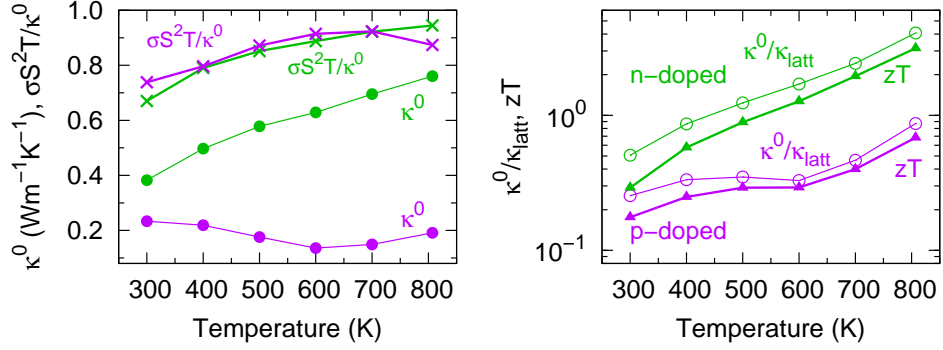}
        \caption{$a)$ Thermal conductivity due to carrier transport
                under isoelectrochemical conditions ($\kappa^{0}$) and the
                ratio $\sigma S^2 T/\kappa^{0}$, $b)$ calculated thermoelectric
                figure of merit ($zT$) and the upper limit of $zT$, namely
                $\kappa^{0}/\kappa_\mathrm{latt}$, as a function of temperature
                for p- and n-doped SnSe.}
        \label{k0}
\end{figure*}

Fig.~\ref{veloc}(b) shows $\Sigma(\epsilon)$ at 300 and 807~K. The difference between the 
magnitudes of $v(\epsilon)$ is largely compensated by the large difference in $\tau_\mathrm{tot}$ 
for p- and n-doped systems at 300~K. However, the difference in $\Sigma(\epsilon)$ 
increases with rising $T$, where the n-doped case presents higher $\Sigma(\epsilon)$ at 807~K. 
These trends are also reflected in the plots of mean-free-paths (mfp), where the mfp of 
n-doped (p-doped) system increases (decreses) with $T$ (see Fig.~S4 of the ESI). Typically 
the mfp increases from 0.5 (300~K) to 1 nm (807~K) for electrons at the CBM in n-doped SnSe.
$\Sigma(\epsilon)$ sets the scale of 
$\sigma$, $PF$ and also $\kappa^0$ (Fig~\ref{k0}), 
and is thus responsible for their larger magnitudes in the n-doped material
as compared to the p-doped material, even at 300 K.

\subsection{High-temperature enhancement of $zT$ in out-of-layer n-doped SnSe}

From Section~\ref{therm} it is clear that the higher $PF$ in n-doped SnSe is the most 
important feature that leads to higher $zT$ in comparison 
to the p-doped case, since $\kappa_\mathrm{tot}$ is very similar for both systems. 
This observation is already clear
in the experimental results of Chang et al.\cite{chang20183d}
However, the high temperature enhancement of $zT$ for n-doped SnSe starting at 600~K 
is a feature that cannot be inferred 
by simple arguments on the basis of the behaviour of the $PF$. The explanation has its roots 
in the decrease of $\kappa_\mathrm{el}$ and the behaviour of $\tau_\mathrm{tot}$, 
as it will be explained below. 

For n-doped SnSe, as shown in Fig.~S5, $PF$/$\kappa_\mathrm{tot}$ 
grows almost linearly with $T$ with a linear coefficient that we call $b$. 
By writing $zT = (PF / \kappa_\mathrm{tot})\times T = (bT)T = bT^2$
we see that $zT$ grows 
approximately quadratically with $T$.
Then, at 600~K, we observe that there is a large increase in $b$, 
so $zT$ begins to grow faster with $T$, 
as evidenced by the quadratic fits for low and high $T$ in Fig~\ref{n_carr}(c). 
However, this change cannot be attributed 
to the $PF$ since it is decreasing (constant) in the range of 400-600~K (600-807~K). 
Consequently, it must be attributed to a faster decrease in $\kappa_\mathrm{tot}$. 

By looking at the plot of $\kappa_\mathrm{tot}$ and $\kappa_\mathrm{latt}$ (Fig.~\ref{veloc}(c)), 
for the n-doped case we observe that $\kappa_\mathrm{latt}$ decreases as $1/T$, without any 
considerable change from 400 to 807~K and hence we can attribute 
the faster decrease in $\kappa_\mathrm{tot}$ to a faster decrease in $\kappa_\mathrm{el}$ 
in the range of 600-807~K (Fig.\ref{therm_prop}(d)). Such faster decrease in $\kappa_\mathrm{el}$ 
also leads $zT_\mathrm{el}$ to increase faster (Fig.\ref{therm_prop}(f)) in 
close agreement with the behaviour of $zT$. This makes sense because
$\kappa_\mathrm{el}$/$\kappa_\mathrm{tot}$ grows almost 
linearly with $T$ (Fig.~\ref{veloc}(d)) and  
$zT$ can be written as $zT = zT_\mathrm{el} \times \kappa_\mathrm{el}/\kappa_\mathrm{tot}$.   

The crucial question is, why does $\kappa_\mathrm{el}$ begin to decrease faster above 600~K? 
Because $\kappa_\mathrm{el}$ is the difference between $\kappa^0$, 
the integral in Eq.\ref{3}, and $PF\times T$, 
both of which are increasing with temperature,
the decay of $\kappa_\mathrm{el}$ with $T$ is a consequence 
of the faster rise of $PF\times T$ as compared to $\kappa^0$. Thus the 
high $PF$ above 600~K is the main cause of the faster decrease in $\kappa_\mathrm{el}$. 
Consequently, the enhancement of $zT$ at higher temperatures 
can be directly connected to the behaviour of $\tau_\mathrm{imp}$, and consequently, $\tau_\mathrm{tot}$, 
since both steadily increase with $T$, maintaining the $PF$ at high values for temperatures above 600~K. 
This constitutes the microscopic origin of the excellent high-$T$ thermoelectric 
performance in out-of-plane n-doped SnSe, since without this feature $zT$ would reach 
a significantly smaller value of around $2.2$ at 807~K, as shown by the 
fit to $zT$ for low temperatures only (Fig.~\ref{n_carr}(c)).

\subsection{Connections to the best thermoelectric}

As pointed out by Mahan and Sofo\cite{mahan1996best}, 
$zT$ has a theoretical upper bound of  $zT_\mathrm{max} =
\kappa^{0}/\kappa_\mathrm{latt}$, which we can calculate directly
using our formalism. 
In fact, several factors impact the magnitude and the behaviour of $\kappa^{0}$ with $T$,  
including: $i)$ the magnitude, temperature dependence, and carrier energy dependence of $\tau_{tot}(\epsilon)$, $v(\epsilon)$ and thus $\Sigma(\epsilon)$; 
$ii)$ the position of the                            
chemical potential, directly related to $n_\mathrm{carr}$, and $iii)$ 
the combined effects of the multiplicative factor, $(\epsilon - \mu)^2$, and
the window function, $\partial f/\partial\epsilon$, that broadens with $T$.
At first glance, it seems that an increase in $\kappa^{0}$ 
would lead to an increase in $\kappa_\mathrm{el}$ and $\kappa_\mathrm{tot}$ and thus a decrease in $zT$.  
However, our results in Fig.~\ref{k0} demonstrate 
the more subtle roles played by $\kappa^{0}$ and $\kappa_\mathrm{el}$ in the determination of $zT$. 

We observe that $\kappa^{0}$ is much higher for n-doped SnSe than p-doped SnSe, clearly indicating that the former has the potential for a higher $zT_\mathrm{max}$, at least
for the specific carrier densities considered in this work.
Additionally, $\kappa^{0}$ steadily increases as a function of $T$ for n-doped SnSe, 
which is related to the fact that its $\tau_\mathrm{tot}$ only increases with $T$. 
For the p-doped case $\kappa^{0}$ decreases with temperature up to 600~K and then start to increase again for $T$ up to 807~K, 
which can be explained by the fact that its $\tau_\mathrm{tot}$ only decreases with $T$. However, 
for temperatures higher than 600~K, the U shaped behaviour of $\tau_{tot}(\epsilon)$ 
allows $\kappa^{0}$ to increase due to the broadening of $\partial f/\partial\epsilon$. 
(A detailed explanation is provided in the ESI accompanying Fig.~S6.) 
The difference between $zT$ and $zT_\mathrm{max}$ 
is inversely proportional to the ratio $\sigma S^2 T/\kappa^{0}$ (see Fig.~\ref{k0}(a)), 
which is intrinsically connected to the magnitude of $\kappa_\mathrm{el}$.  
In fact, $\kappa_\mathrm{el}$ is smaller when $\sigma S^2 T/\kappa^{0}$ is larger, and 
$\kappa_\mathrm{el} \to 0$ as $\sigma S^2 T/\kappa^{0} \to 1$. Thus, our results clearly show 
that 
$zT$ approaches the upper limit, $\kappa^{0}/\kappa_\mathrm{latt}$, only when 
$\kappa_\mathrm{el}$ is small. 
In view of this picture, the use of band-pass energy filters\cite{mckinney2017search} 
over transport distribution functions 
is not always productive, since it cuts off $\kappa^0$ in order 
to reduce $\kappa_\mathrm{el}$ and $\Lambda$,
with the disadvantage of lowering the upper limit.  

\section{Conclusions}

We investigated the reasons behind the excellent TE performance in out-of-plane n-doped SnSe
by employing dual interpolation first-principles calculations of non-polar 
and screened polar e-p coupling combined with a semi-empirical methodology to compute the scattering of 
charge carriers by ionized impurities. 
Using reported values for $S$ and $\sigma$ to self-consistently determine the carrier density and ionized impurity concentration, 
we calculated the TE transport properties of SnSe, including $\kappa^0$, $\kappa_\mathrm{el}$, $\Lambda$, 
$zT_{el}$ and $zT$, for both n- and p-doping and temperatures up to 807~K. 
Our calculations predict an ultrahigh-$zT = 3.1$ for n-doped SnSe at 807~K.   
In order to understand the high $zT$ as well as the 
enhancement of $zT$ above 600~K for n-doped SnSe, we analyzed 
several important microscopic quantities that jointly impact the overall transport properties, 
such as the average electronic group velocities as well as the carrier energy and temperature 
dependence of scattering mechanisms and their respective relaxation times. 

Our results show that the scattering by ionized impurities is 
the dominant scattering mechanism in n-doped SnSe up to 700~K,
while non-polar phonon scattering dominates for higher temperatures. 
In the p-doped case, these two mechanisms are comparable throughout the 
temperature range, but have different dependence on the carrier energy.
All the RTs calculated show a decrease with temperature, \textit{except} 
for $\tau_\mathrm{imp}$ in n-doped SnSe. Because impurity scattering 
is dominant for n-doping, $\tau_\mathrm{tot}$ increases with temperature, 
even after the crossover to non-polar e-p scattering at 700~K. 
This behaviour of $\tau_\mathrm{tot}$ that increases with temperature, 
in conjunction with the intrinsically higher electronic group velocities in n-doped SnSe, 
act cooperatively to produce a high $PF$ and high $\kappa^{0}$ and,
simultaneously, to reduce $\kappa_\mathrm{el}$ even faster beyond 600~K, 
allowing for the ultrahigh-$zT = 3.1$ at 807~K.
Note that $\kappa_\mathrm{latt}$ 
is smaller in n-doped SnSe, possibly due to the
heavier doping of the sample. This contributes to the increase in $zT$, 
though it is not the decisive factor.

For p-doped SnSe, the intrinsically lower electronic group velocities, along with
$\tau_\mathrm{tot}$ that decreases with $T$, makes the $PF$ and $\kappa^0$ much smaller than in the n-doped case, and
with the tendency to decrease further with rising $T$. At higher temperatures,
some hope arises for the p-doped case due to the
U shaped behaviour of $\tau_\mathrm{tot}(\epsilon)$ that acts to increase $\kappa^0$ and thereby
increase the upper limit ($\kappa^{0}/\kappa_\mathrm{latt}$) of $zT$.
At 807~K, the increase in
$n_\mathrm{carr}$ due to the formation of Sn vacancies tends to
increase the $PF$, but it is counteracted by the reduced $\tau_\mathrm{tot}$.
Consequently, with low $PF$, $\kappa_\mathrm{el}$ increases as $PF \times T/\kappa^{0}$ decreases (see Fig.~\ref{k0}(a)), causing 
the calculated $zT$ to move further away from the upper limit, $\kappa^{0}/\kappa_\mathrm{latt}$.
The plots of $zT$ and $zT_\mathrm{max}$ for p-doped SnSe in Fig.\ref{k0}(b) clearly demonstrate that $zT$ 
approaches its upper limit when $\kappa_\mathrm{el}$ is very small.
Informed by our results, optimization of the TE performance of p- and n-doped SnSe can be 
achieved by adjusting $n_\mathrm{carr}$ in order to optimize $\kappa^0$. This is the subject for future
work.

\setcounter{secnumdepth}{0}

\vspace*{7 mm}

\section*{Electronic Supplementary Information}
Extra data and analyses are provided in the electronic supplementary information.

\section*{Conflicts of interest}
There are no conflicts to declare.

\section*{Acknowledgements}

ASC and AA gratefully acknowledge support from the Brazilian agencies CNPq and FAPESP under
Grants \#2010/16970-0, \#2013/08293-7, \#2015/26434-2, \#2016/23891-6, \#2017/26105-4, and \#2019/26088-8.
DTL is supported by NSF DMREF Grant No. 1922165 and by DOE Basic Energy Science Award No. DE-SC0019300, and is a participant in the NSF
STC Center for Integrated Quantum Materials, NSF Grant No. DMR-1231319.
The calculations were performed at CCJDR-IFGW-UNICAMP in Brazil.

\vspace*{7 mm}
\section*{Computer Code Availability}
All computer implementations of the methodology developed in this project were written
in Fortran 90 and are available upon request.


\bibliographystyle{apsrev4-1}
{\footnotesize
\bibliography{Ref2.bib}}

\end{document}



\title{Electronic Supplementary Information: Microscopic origin of the excellent thermoelectric performance in n-doped SnSe}

\author{Anderson S. Chaves}
\affiliation{Gleb Wataghin Institute of Physics, University of Campinas, PO Box 13083-859, Campinas, SP, Brazil}
\author{Daniel T. Larson}
\affiliation{Department of Physics, Harvard University, Cambridge, Massachusetts, 02138, USA}
\author{Efthimios Kaxiras}
\affiliation{Department of Physics, Harvard University, Cambridge, Massachusetts, 02138, USA}
\affiliation{John A. Paulson School of Engineering and Applied Sciences, Harvard University, Cambridge, Massachusetts, 02138, USA}
\author{Alex Antonelli}
\affiliation{Gleb Wataghin Institute of Physics and Center for Computing in
Engineering \& Sciences, University of Campinas, PO Box 13083-859, Campinas, SP, Brazil}

\date{\today}


\maketitle


\section{Increase of $\kappa^0$ above 600~K for p-doped SnSe}

In order to explain the increase of $\kappa^0$ above 600~K for p-doped SnSe,
we resort to some deeper analysis of the 
energy moment within the Boltzmann Transport Equation given
by the integrand to calculate $\kappa^0$ (see the main text), which is defined as
\begin{equation}
\label{klinha}
\kappa^{\prime} \propto \Sigma (\epsilon) (\epsilon - \mu)^2 \left(-\frac{\partial f}{\partial \epsilon}\right)~,
\end{equation}
which includes the window function $\partial f/\partial\epsilon$, and
\begin{equation}
\label{k2linhas}
\kappa^{\prime \prime} \propto \Sigma (\epsilon) (\epsilon - \mu)^2~,
\end{equation}
which does not include it. The results for $\kappa^{\prime}$ and $\kappa^{\prime \prime}$ at 700 and 807~K
for both p- and n-doped SnSe are shown in Fig.~\ref{k0}.
In order to display these results in the same plot, we placed both, VBM and CBM, at 0~eV and reversed the
sign of the energy scale for the p-type material.
For p-type SnSe, we clearly observe an abrupt change in the behaviour of $\kappa^{\prime}$ and $\kappa^{\prime \prime}$ for
both temperatures, 700~K and 807~K, around 0.3~eV as indicated by green arrows in Fig.~\ref{k0}.
This feature is not observed
in the case of n-type SnSe. Clearly, this abrupt change may be correlated with
the U shaped behaviour of $\tau_\mathrm{npol}$ that is the dominant relaxation time for the p-type material
in this energy regime (see the main text for a proper explanation). 
In fact, the increase of $\tau_\mathrm{npol}$ effectively
enhances the values of $\kappa^{\prime}$ and $\kappa^{\prime \prime}$
through the kernel $\Sigma$, for energies higher than 0.3~eV, that means in this case lower than -0.3~eV below the VBM.
It is clear from Fig.~\ref{k0}(a) that the change in the behaviour of $\kappa^{\prime}$ results in an increase of
its integral over energy ($\kappa^0$).
In contrast, for n-type SnSe, the smooth behaviour of
$\tau_\mathrm{imp}$ and $\tau_\mathrm{npol}$ is reflected in the data displayed in Fig.~\ref{k0}, in which
no abrupt change in the behaviour of $\kappa^{\prime}$ and $\kappa^{\prime \prime}$ is observed.
The role of the temperature dependent window function, $\partial f/\partial\epsilon$,
is to change the sampling of energy points used in the
integration for the calculation of $\kappa^0$,
that is, it broadens with rising $T$.
Additionally, the multiplicative term $(\epsilon - \mu)^2$ increases
the importance of energy values away from the chemical potential, $\mu$, that lies
close to the band edge. This term enhances the importance of energy values higher than 0.3~eV
in the sampling. Therefore, the increase of $\kappa^0$
in the case of the p-type material from 600 to 807~K is a consequence of threefold effects: $i)$
the abrupt change caused by the behaviour of $\tau_\mathrm{npol}$ through $\Sigma$,
associated with both $ii)$ the broadening in the window function caused by temperature effects
and $iii)$ the sampling of higher energy values in energy integration.

\renewcommand{\thefigure}{S1}
\begin{figure*}
\centering
\includegraphics[width=1\textwidth]{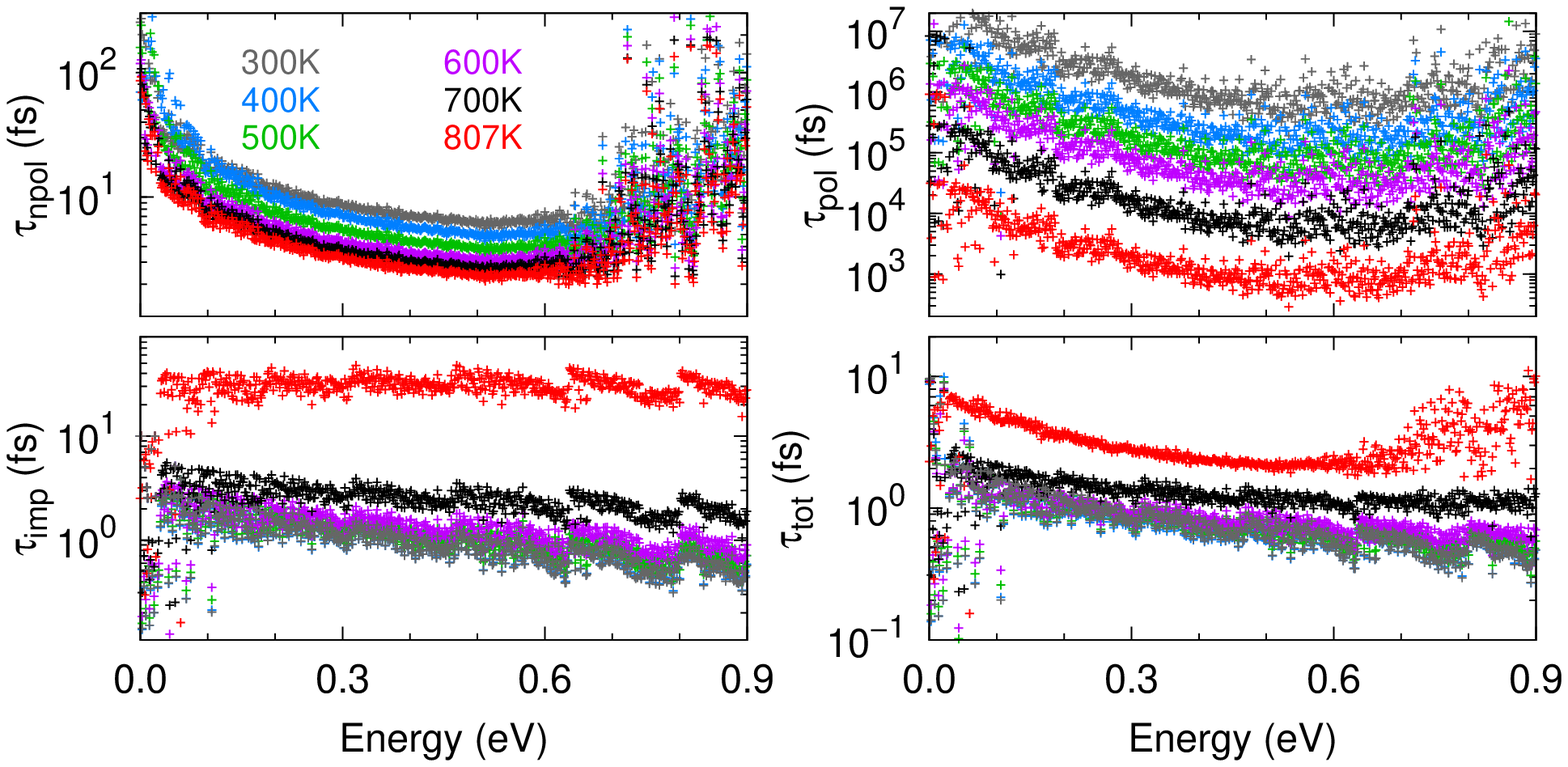}
\vspace*{6mm}\caption{\label{figureS2}
Temperature dependence of each scattering mechanism as a function of carrier energy, namely, the nonpolar scattering by phonons ($\tau_{npol}$), 
screened polar scattering by optical phonons ($\tau_{pol}$), screened scattering by ionized impurities ($\tau_{imp}$) 
and total scattering ($\tau_{tot}$), in n-doped SnSe. }
\end{figure*}

\renewcommand{\thefigure}{S2}
\begin{figure*}
\centering
\includegraphics[width=1\textwidth]{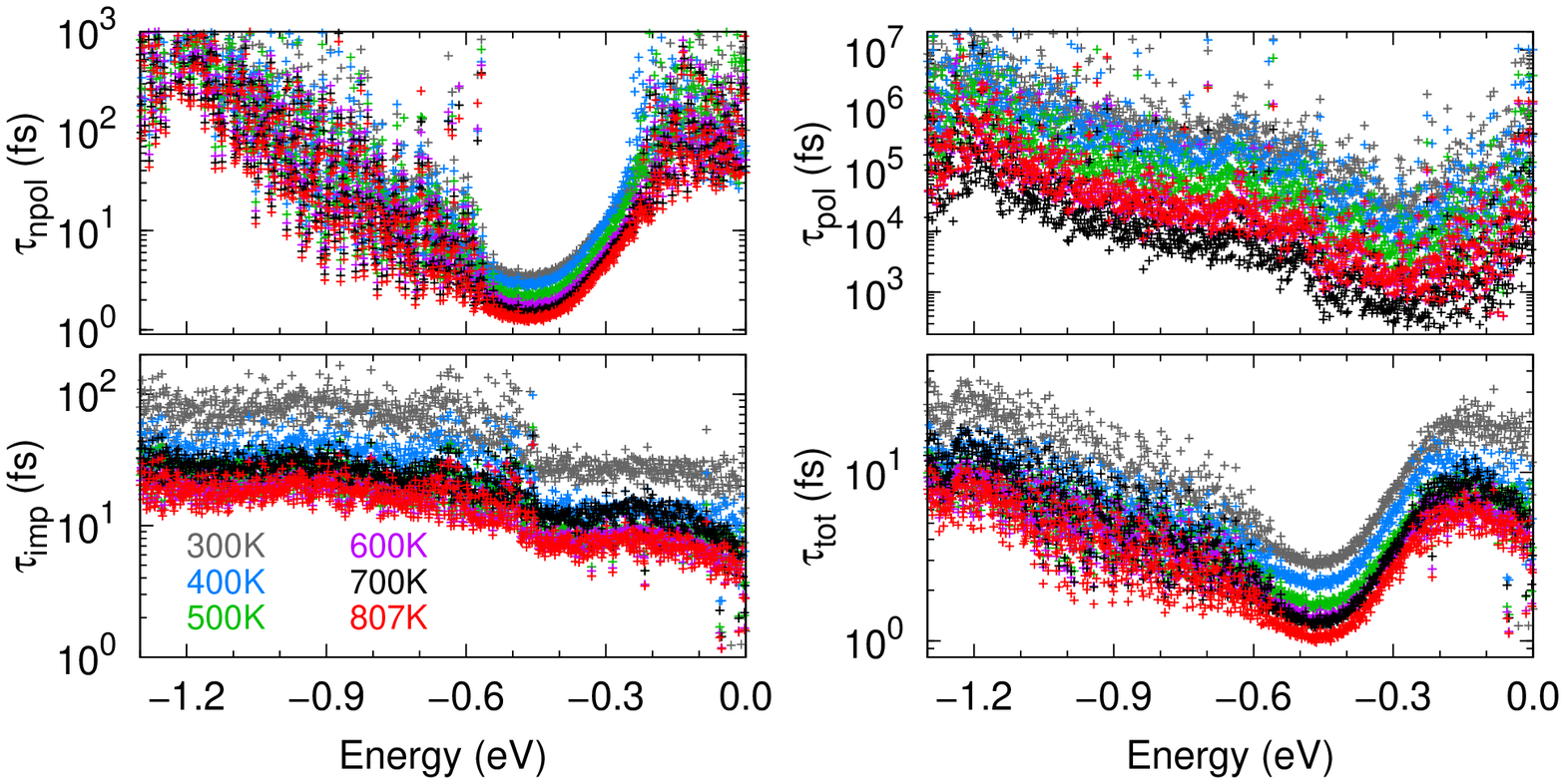}
\vspace*{6mm}\caption{\label{figureS3}
Temperature dependence of each scattering mechanism as a function of carrier energy, namely, the nonpolar scattering by phonons ($\tau_{npol}$),
screened polar scattering by optical phonons ($\tau_{pol}$), screened scattering by ionized impurities ($\tau_{imp}$) 
and total scattering ($\tau_{tot}$), in p-doped SnSe.}
\end{figure*}

\renewcommand{\thefigure}{S3}
\begin{figure*}	
	\centering
	\includegraphics[width=1\textwidth]{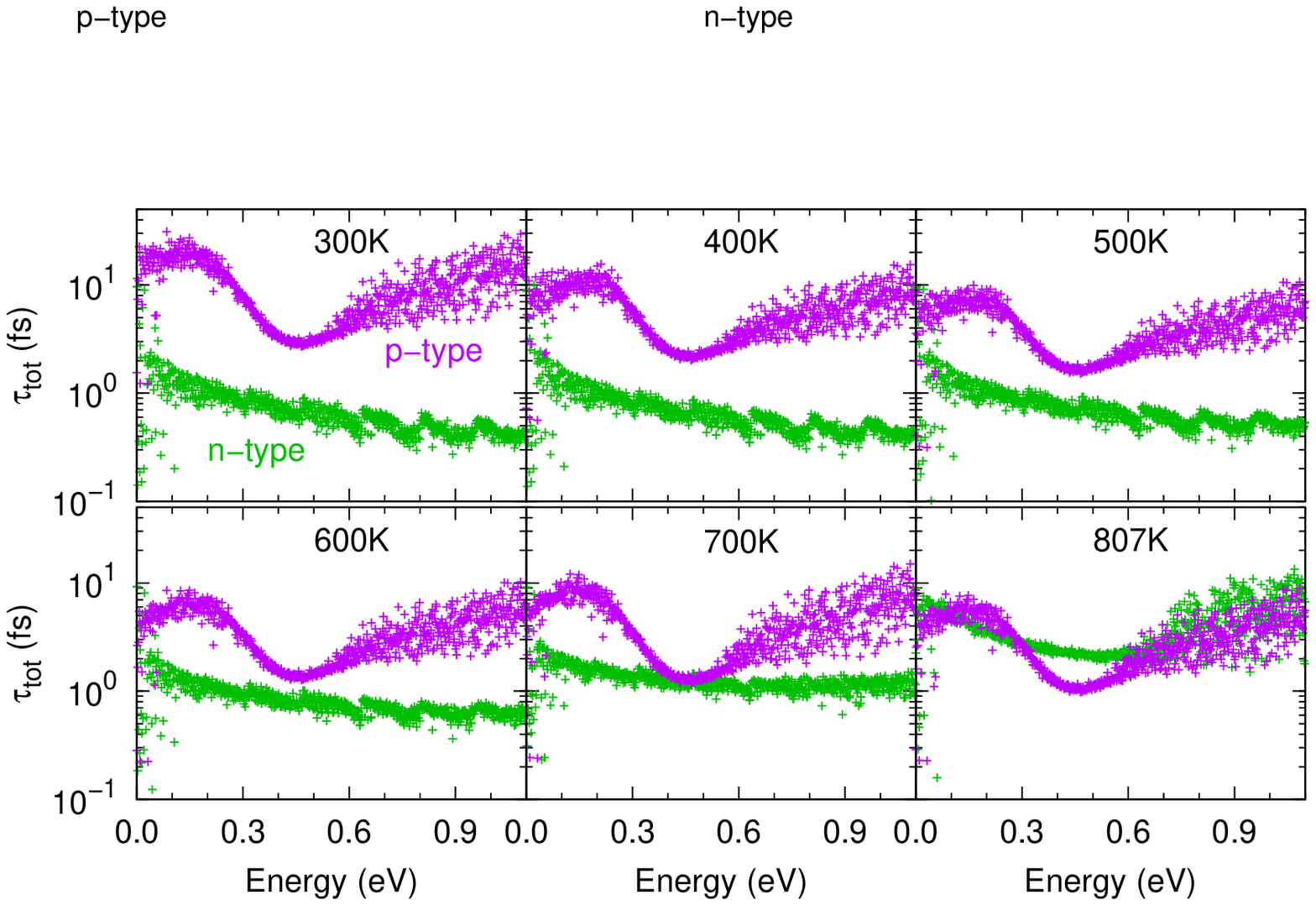}
	\vspace*{6mm}\caption{\label{figureS1}
		Comparison between the total relaxation time, $\tau_{tot}$, as a function of carrier energy and temperature (300-807K) 
		for n-doped and p-doped SnSe in the out-of-layer direction with similar carrier densities. 
		$\tau_{tot}$ is given in terms of nonpolar, screened polar and screened scattering by ionized impurities on the basis of 
		Mathiessen's rule, as explained in the main text.}
\end{figure*}

\renewcommand{\thefigure}{S4}
\begin{figure*}
\centering
\includegraphics[width=1\textwidth]{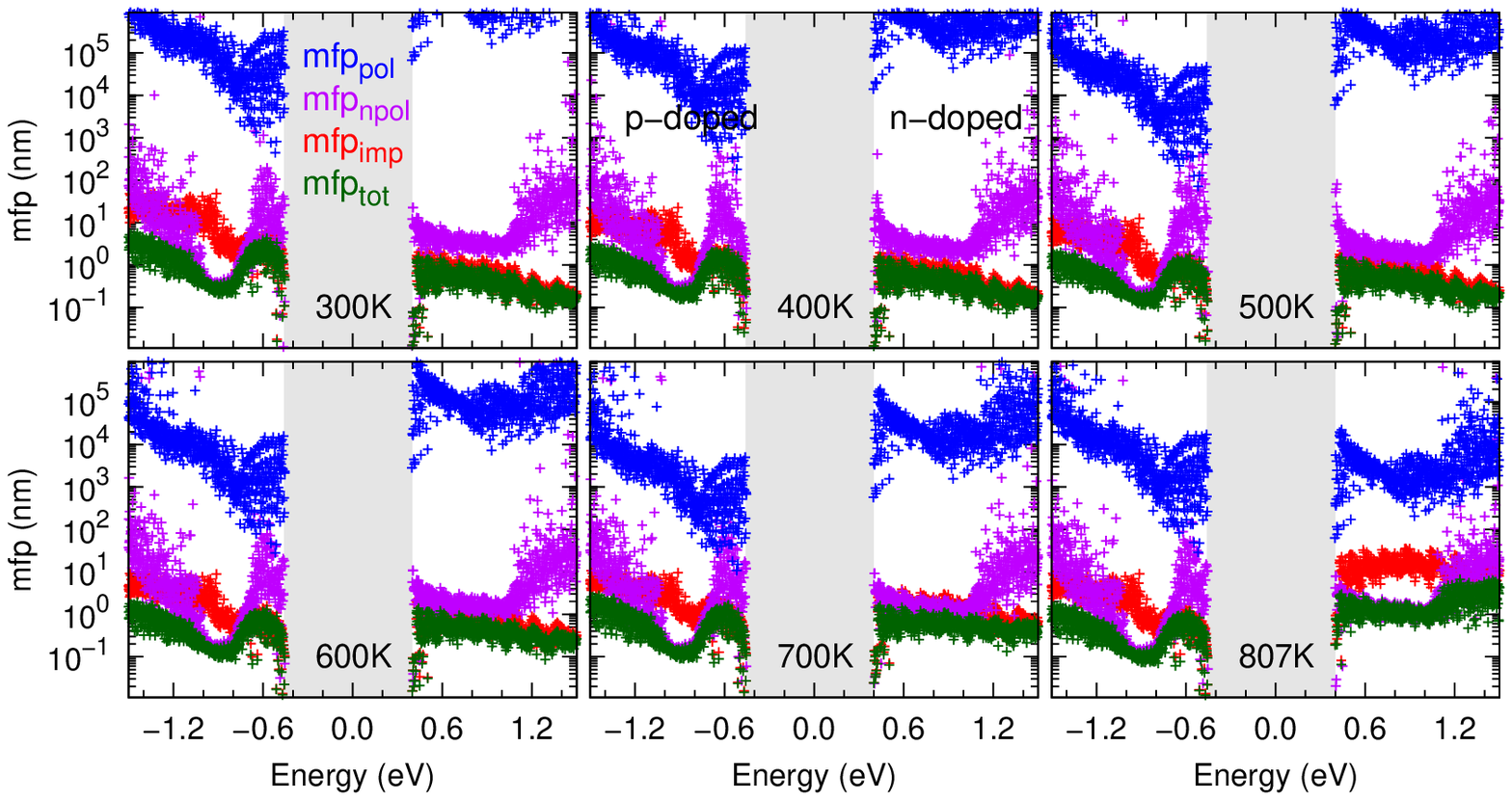}
\vspace*{6mm}\caption{\label{figureS4}
Temperature dependence (300-807K) of the mean free path (mfp) as a function of carrier energy
calculated in the out-of-layer direction of p- and n-doped SnSe 
as $mfp (\epsilon) = \sqrt{\frac{\sum_{n,{\bf{k}}} \tau_{n,{\bf{k}}} v_{n,{\bf{k}}} \tau_{n,{\bf{k}}} v_{n,{\bf{k}}} \delta (\epsilon-\epsilon_{n,{\bf{k}}})}{\sum_{n,{\bf{k}}} \delta (\epsilon-\epsilon_{n,{\bf{k}}})}}$
for each scattering mechanism, namely, nonpolar scattering of acoustic and optical phonons ($\tau_{npol}$),
scattering by ionized impurities ($\tau_{imp}$) and screened polar scattering of optical phonons ($\tau_{pol}$).
The total mfp due to the total relaxation time ($\tau_{tot}$), which is calculated from the Mathiessen's rule,
is also shown as a function of the carrier energy and temperature.
The light grey rectangles represent the gap region with an energy gap of 0.86eV.}
\end{figure*}

\renewcommand{\thefigure}{S5}
\begin{figure*}[t]
\centering
\includegraphics[width=0.46\textwidth]{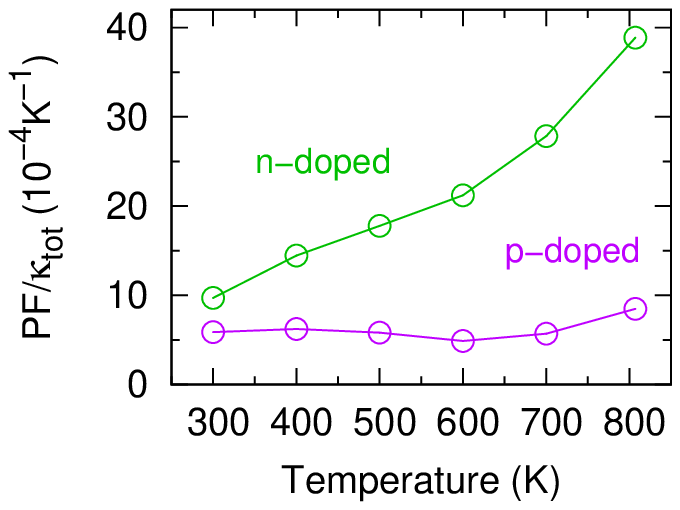}
\caption{Ratio between the power factor ($PF = \sigma S^2$) and total thermal conductivity ($\kappa_\mathrm{tot}$) as a function of temperature (300-807~K) for p- and n-doped SnSe.}
\label{PFoverktot}
\end{figure*}

\renewcommand{\thefigure}{S6}
\begin{figure*}[t]
\centering
\includegraphics[width=0.4\textwidth]{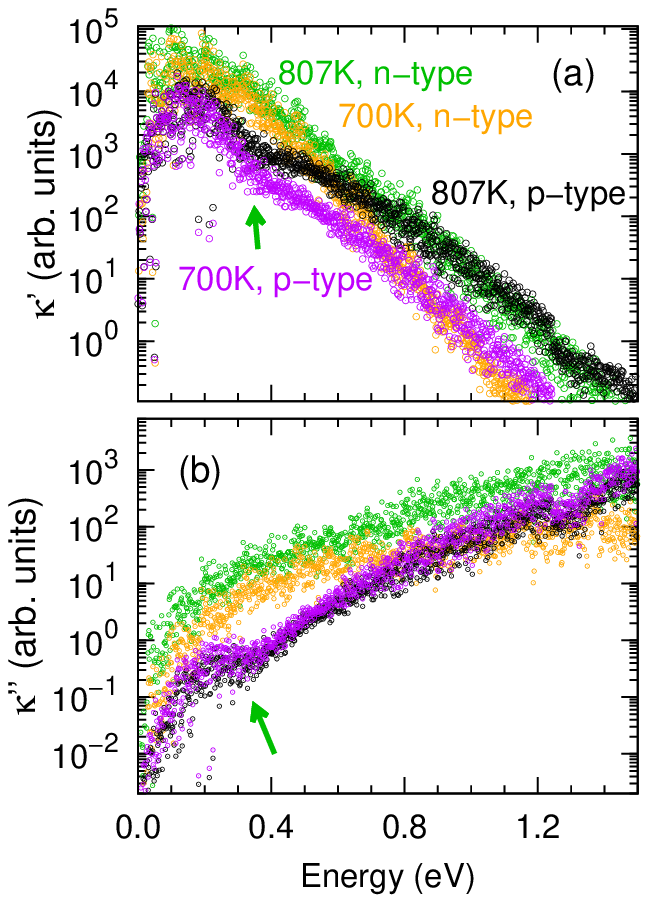}
\caption{Energy moments of BTE: (a) $\kappa^{\prime}$ (Eq.\ref{klinha}, integrand to calculate $\kappa^0$) 
         and (b) $\kappa^{\prime \prime}$ (Eq.\ref{k2linhas})
         for 700 and 807~K for p- and n-doped SnSe in function of the carrier energies. The values were shifted
         in order to place both VBM and CBM at 0~eV.}
\label{k0}
\end{figure*}
